# Nanoantennas Design for THz Communication: Material Selection and Performance Enhancement

Sasmita Dash
Computer Science Department
University of Cyprus
Nicosia, Cyprus
sasmitadash30@ieee.org

Christos Liaskos
Institute of Computer Science
Foundation of Research and
Technology Hellas
Heraklion, Greece
cliaskos@ics.forth.gr

Ian F. Akyildiz
School of Electrical and Computer
Engineering
Georgia Institute of Technology
Atlanta, USA
ian@ece.gatech.edu

Andreas Pitsillides
Computer Science Department
University of Cyprus
Nicosia, Cyprus
andreas.pitsillides@ucy.ac.cy

## ABSTRACT

In the development of terahertz (THz) communication systems, the nanoantenna is the most significant component. Especially, the focus is to design highly directive antennas, because it enhances the performance of the overall system by compensating the large path loss at THz and thus improves the signal-to-noise ratio. This paper presents suitable material for nanoantenna design and the advancement in their performance for THz communications. Copper, Graphene, and carbon nanotube materials are used as promising candidates for nanoantenna design. The performance of nanoantennas is carried out by analyzing the properties and behavior of the material at THz. Results show that the Graphene nanoantenna provides better performance in terms of miniaturization, directivity, and radiation efficiency. Further, the performance enhancement of the nanoantenna at THz is studied by dynamically adjusting the surface conductivity via the chemical potential of Graphene using the electric field effect. The performance of the nanoantenna is enhanced in terms of high miniaturization, high directivity, low reflection, frequency reconfiguration, and stable impedance. The THz nanoantennas using Graphene have the potential to be used for THz communication systems. In view of the smart THz wireless environment; this paper finally presents a THz Hypersurface using Graphene meta-atoms. The user-side Graphene nanoantennas and environment-side Graphene Hypersurface can build a promising smart THz wireless environment.

## KEYWORDS

Nanoantenna, Terahertz, Graphene, Carbon nanotube, Copper, Wireless communication, Hypersurface.

## 1 Introduction

The bandwidth requirements for wireless communications have increased rapidly over the last few decades. In order to tackle the situation, advanced modulation techniques have been applied to increase the spectral utilization efficiency [1]. This not only has increased the data rates to a large extent, but also, enhanced the frequency reuse within a volume of space. However, the channel capacity upper limit is restricted by Shannon's formula, even with the use of multi-input multi-output (MIMO) techniques. Therefore, the only way to provide sufficient transmission capacity is by accessing higher carrier frequencies transmission bands. This desire for higher carrier frequency or more bandwidth led the researchers to take advantage of the terahertz (THz) spectrum, that is, electromagnetic waves with frequencies ranging from 0.1 to 10 THz [2]. Besides this intrinsic advantage of high bandwidth, THz wireless communication has other advantages when compared with either microwave link or infrared (IR) based systems, such as (i) more directional than microwave/millimeter links, (ii) security, (iii) low attenuation compared to IR, (iv) smaller scintillation effects compared to IR, *etc*. [3]. Because of these advantages, THz technology has grown dramatically over the last two decades and found its application in various fields. But, even today, these applications are not fully implemented due to immature state of THz technology in terms of sources, detectors, antennas, and other basic components capable of working effectively in this frequency range.

In the last few years, several metal nanoantennas and array structures including photoconductive dipole antennas, planar antennas, bow-tie antennas, horn antennas, reflector antennas, and lens antennas have been designed for THz applications [4-9]. The implementation of metal nanoantenna design in THz frequency remains a challenge. The dimension of the traditional metal nanoantenna in this frequency range is in the order of micrometers. There lies a technological challenge in micro fabrications. Furthermore, at higher frequencies, the conductivity and skin depth of conventional metals used for nanoantenna design (e.g., copper, gold, silver, *etc*.) decreases, leading to degradation of radiation efficiency and high propagation losses.

The other option is the use of carbon-based nanomaterials (e.g., Graphene and carbon nanotube) for THz nanoantenna designs [10-16] [17-27]. This paper compares Graphene, carbon nanotube and copper material properties at THz and presents the most promising material for THz nanoantenna design and the enhancement in their performances. A discussion on Graphene Hypersurface (HSF) for the smart THz wireless environment is also provided in this paper.

## 2 THz Communication

The THz band in the electromagnetic spectrum lies between the microwave and infrared frequencies. The THz electromagnetic spectrum has several benefits for wireless communications [3], [28]. The achievement of the data rates of 10 Gbps is comparatively difficult in the microwave band due to the narrow bandwidth. The data rate of 10–100 Gbps is realized by raising the carrier frequencies at 100–500 GHz [29]. The opportunity for large bandwidth in the THz band leads to the possibility of easy high data rate transmission [30]. THz communication is promising for wireless communications systems, particularly for the short-range indoor environment. The use of THz frequency allows for miniaturized antennas, which enables massive MIMO techniques for enhancement of spectral efficiency and directivity. In spite of the advantages, THz communication has a few limitations. The discussion for the THz channel model is found in the literature, e.g., [31]. In the THz band, the free space path loss is more significant than at lower frequencies. This is the main reason to have less received power than the transmitted power. Furthermore, the THz signal suffers from both molecular absorption loss and spreading loss. In the THz band, the total path loss is

$$A(f,d) = A_{ma}(f,d) + A_s(f,d) \quad (1)$$

where $A_{ma}(f,d)$ and $A_s(f,d)$ are the molecular absorption loss and spreading loss at frequency $f$ and distance $d$. On account of the molecular absorption loss, several high attenuation levels are defined [31]. In the THz band, the spreading loss is 60 dB higher compared to the microwave band. Therefore, the THz band is merely appropriate for short-range communications where the range is in the order of a few tens of meters.

## 3 Materials for THz Nanoantenna

Nanoantennas for THz communication have been reported in the last few years [18-20], [23-25]. The research focus is towards the design of highly directive nanoantennas because they enhance the performance of the overall system by compensating for the large path loss at THz and thus improving the signal-to-noise ratio. Furthermore, broad bandwidth and narrow beam THz nanoantennas provide high range and spatial resolution, respectively. In the development of THz communication systems, the nanoantenna is the most significant component. The materials so far used for nanoantenna design in the THz band are Graphene, carbon nanotube and conventional metal copper [4-27]. However, the selection of suitable material for nanoantenna design is important for THz communication. We present suitable materials for THz nanoantennas in the coming sections.

### 3.1 Graphene

The latest addition to the family of carbon allotropes is Graphene, two-dimensional $sp^2$-bonded carbon atoms packed in a honeycomb lattice [32]. Graphene has extraordinary electronic properties at THz, such as high electron mobility of $2\times10^5$ cm$^2$V$^{-1}$s$^{-1}$ and high current density of $10^9$ A/cm [33], [34]. The unique properties of Graphene at THz is the surface plasmon polariton (SPP) wave propagation [35]. Due to the two-dimensional nature, Graphene surface plasmons exhibit strong confinement, low losses and high tunability. Graphene SPP wave propagates with SPP wavelength $\lambda_{SPP}$ much less than the wavelength of free-space $\lambda_0$ (i.e., $\lambda_{SPP} \ll \lambda_0$) [36]. Owing to the SPP properties at THz range, Graphene enables plasmonic nanoantennas in the THz frequency regime.

According to Kubo formalism, the surface conductivity $\sigma_s$ of Graphene is given as [37]

$$\sigma_s = \sigma_{intra} + \sigma_{inter}$$

where $\sigma_{intra}$ and $\sigma_{inter}$ are the conductivity due to intraband and interband transition respectively. The intraband conductivity dominates in the THz band, given by

$$\sigma_s(\omega,\mu_c,\tau,T) = -j\frac{e^2\kappa_B T}{\pi\hbar^2(\omega-j\tau^{-1})}\left[\frac{\mu_c}{\kappa_B T} + 2\ln\left(e^{-\mu_c/\kappa_B T}+1\right)\right] \quad (2)$$

where $e$ is the charge of electron, $\omega$ is the angular frequency, $k_B$ is the Boltzmann's constant, $\mu_c$ is the chemical potential, $T$ is the temperature, $\tau$ is the relaxation time, and $\hbar$ is the reduced Planck's constant.

Graphene surface impedance can be calculated as $Z_S = 1/\sigma_s = R_S(V_b) + jX_S(V_b)$, where $V_b$ is the bias voltage. The surface impedance and surface conductivity of Graphene can be dynamically controlled via chemical potential by applying a DC bias voltage. Graphene surface impedance exhibits highly inductive nature at THz.

### 3.2 Carbon Nanotube

Carbon nanotube (CNT) is formed by rolling of Graphene sheet [38]. The rolling along the x-axis and y-axis form zigzag CNT and armchair CNT respectively. CNT is also categorized as single-walled CNT or multiwalled CNT. CNT has high electron mobility of $8\times10^4$ cm$^2$V$^{-1}$s$^{-1}$, and high current density of $10^9$ A/cm [39]. CNT supports plasmonic wave propagation at THz frequency. Because of the curvature effect, CNT has more plasmonic losses and less tunability behavior in comparison to Graphene.

The electronic behavior of the armchair and zigzag CNT variations are different. Zigzag CNTs exhibit both semiconducting and metallic behavior, whereas armchairs CNTs show only metallic behavior. The conductivity of CNT consists of both intraband and interband contribution. Conductivity due to

Intraband transition is more significant in the THz frequency regime. The intraband conductivity of CNT $\sigma_{CNT}$ can be expressed as [12]

$$\sigma_{CNT} = -j\frac{2e^2 v_f}{\pi^2 \hbar r(\omega - j\tau^{-1})} \quad (3)$$

where $v_f$ is Fermi velocity, $e$ is the electronic charge, $\hbar$ is the reduced Plank's constant, $r$ is the radius of CNT, $\omega$ is the angular frequency, and $\tau$ is the electron relaxation time. Here, we assume armchair CNT. The surface impedance of CNT can be calculated as

$$Z_{CNT} = \frac{1}{2\pi r \sigma_{CNT}} = \frac{j\pi\hbar(\omega - j\tau^{-1})}{4e^2 v_f} = \frac{\pi\hbar\tau^{-1}}{4e^2 v_f} + j\omega\frac{\pi\hbar}{4e^2 v_f} \quad (4)$$

From the above equation, CNT scattering resistance $R_{CNT}$ and kinetic inductance $L_{CNT}$ can be written as,

$$R_{CNT} = \frac{\pi\hbar\tau^{-1}}{4e^2 v_f}, \text{ and } L_{CNT} = \tau R_{CNT} = \frac{\pi\hbar}{4e^2 v_f} \quad (5)$$

The scattering resistance and kinetic inductance of CNT surface impedance are of the same order in the THz range. Scattering resistance of CNT depends on relaxation time and independent of frequency. The kinetic inductance of CNT increases with frequency. Hence, CNT scattering resistance remains same, whereas the kinetic inductance increases with frequency in the THz frequency regime.

### 3.3 Copper

It is widely known that copper is an excellent electrical conductor and the most commonly used metal at RF and microwave frequency regimes. Copper has an electron mobility of 32 cm$^2$/Vs, and a current density of $10^6$ A/cm.

At THz frequency regime, the conductivity and impedance of copper using the Drude theory can be expressed, as [40]

$$\sigma_D^{cu} = \frac{ne^2\tau}{m(1+j\omega\tau)} = \frac{\sigma_0^{cu}}{1+j\omega\tau} \text{ and } Z_{cu} = \sqrt{\frac{j\omega\mu_0}{\sigma_D^{cu} + j\omega\varepsilon_0}} \quad (6)$$

where $\sigma_0^{cu} = (ne^2\tau/m)$ is the dc-conductivity, $n$ is the conduction electron density, and $m$ is the mass of the electron. The surface impedance of copper at THz frequency can be expressed as,

$$Z_{cu} = R_{cu} + j\omega\left(L_{cu}^i + L_{cu}^k\right) \quad (7)$$

where $L_{cu}^i$ and $L_{cu}^k$ are the internal and kinetic inductance respectively, and $R_{cu}$ is the ohmic resistance. In the case of copper, kinetic inductance is larger than its internal inductance and smaller than the ohmic resistance at low THz frequencies. Although copper is a good electrical conductor at microwave frequencies, it does not exhibit the same behavior at THz. The skin depth and conductivity of copper at THz frequencies are less than at microwave frequencies. Thus, the design of an effective copper THz nanoantenna is difficult. The decline of skin depth and conductivity of copper at THz frequency leads to high propagation losses and degradation of the radiation efficiency.

## 4 Performance Analyses of Nanoantennas at THz

In order to have a sharp comparison between the copper, Graphene, and CNT nanoantennas at THz, the analysis has been studied in two phases. In the first phase, the size of the nanoantenna was kept constant, whereas, in the second phase, the frequency was kept constant. The motivation was to find the best material that can be used for the nanoantenna design with superior performance at THz band.

In the first phase, nanoantennas of the same size at low THz frequency regime are considered for the comparison of the antenna performance for copper, Graphene, and CNT materials [18]. Fig. 1 shows the schematic of Graphene, CNT and copper nanoantenna of the same length. The nanoantennas are placed over SiO$_2$ substrate in all three cases. Here, FEM based electromagnetic solver is used to validate the nanoantennas. Copper, Graphene and CNT nanoantenna of the same length 71 μm resonate at 1.90 THz, 0.81 THz, and 1.42 THz respectively. The result showed that the resonant frequency of Graphene nanoantenna remains at lower frequency than copper and CNT nanoantennas. It has also been found that the directivity and the radiation efficiency of Graphene nanoantenna are higher than copper and CNT nanoantenna. The directivity of nanoantennas for copper, Graphene and CNT material are 2.2 dBi, 4.5 dBi, and 3.5 dBi respectively. Further, we can notice that the directivity of the CNT nanoantenna is higher than that of copper nanoantenna at low THz frequency range.

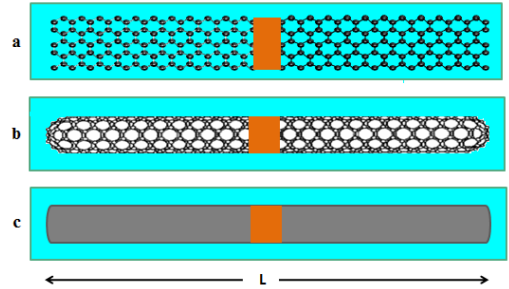

**Figure 1:** Top view of Nanoantennas of the same length. (a) Graphene, (b) CNT, and (c) Copper.

In the second phase, the performances of copper, Graphene, and CNT nanoantennas at same resonant frequency of 1 THz are numerically analyzed and compared to find the most suitable material for THz nanoantenna [17]. It has been found that, at 1 THz frequency, length of copper nanoantenna is 139 μm (=$\lambda_0$/3), whereas the length of Graphene and CNT nanoantennas are 68 μm (=$\lambda_0$/4.4) and 99 μm (=$\lambda_0$/2) respectively. This reveals that the Graphene nanoantenna favors high miniaturization over CNT and copper nanoantennas. This analysis also shows that the Graphene nanoantenna exhibits maximum directivity. The performance of

Graphene, CNT, and copper nanoantenna at 1 THz is summarized in Table.1.

The above analysis reveals that (i) Graphene is highly inductive and characterized by SPP at THz. Owing to excellent electronic properties and the propagation of TM SPP waves at THz band, Graphene nanoantennas exhibit higher directivity and higher miniaturization than copper and CNT nanoantennas. (ii) The kinetic inductance of copper is much less than the ohmic resistance and CNT has large kinetic inductance at THz band. Due to the kinetic inductive effect, CNT supports slow-wave propagation. (iii) At THz, owing to the support of slow-wave propagation and larger conductivity than copper, CNT nanoantennas provide high miniaturization and larger directivity than copper nanoantennas.

**Table 1**: Performance of Graphene, CNT and Copper nanoantennas

| Nanoantenna (@ 1 THz) | Graphene | CNT | Copper |
|---|---|---|---|
| Length (μm) | $\lambda_0/4.4$ | $\lambda_0/3$ | $\lambda_0/2$ |
| Directivity (dBi) | 4.3 | 3.0 | 2.2 |

## 5 Performance Enhancement in Nanoantenna Design at THz

After establishing Graphene as the most suitable material for designing THz nanoantennas in the previous section, this section focuses on the advancement and performance enhancement of Graphene nanoantennas. As far as antenna application of Graphene is concerned, the important property is its ability to support the propagation of SPP waves and because of which, the working frequency of Graphene nanoantenna remains in the THz band.

The schematic of the Graphene nanoantenna is shown in Fig. 2. The performance of Graphene nanoantenna is enhanced by dynamically adjusting the Graphene conductivity using electric field effect. The performance enhancement of the Graphene nanoantenna is in terms of high directivity, low reflection, stable impedance, high miniaturization and frequency reconfiguration [20].

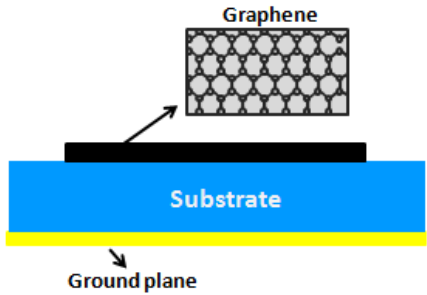

Figure 2: Schematic of Graphene nanoantenna

In order to understand the reconfiguration performance of Graphene nanoantenna at THz, one needs to know the conductivity property of Graphene. In the THz frequency regime, the intraband conductivity is the dominant contribution and shows the dependence on chemical potential $\mu_c$. This indicates that the tunability of Graphene conductivity can be achieved by adjusting the chemical potential of Graphene. The application of an external voltage at Graphene controls the chemical potential. In Graphene nanoantenna over $SiO_2/Si$ substrate, the $SiO_2$ layer provides a convenient way to control dynamically the Graphene conductivity by applying an external gate voltage $V_g$ between the Graphene and the silicon layer [20]. This electric field effect by means of external gate voltage at the Graphene layer controls the electromagnetic properties of Graphene.

A closed-form equation for chemical potential $\mu_c$ and the gate voltage $V_g$ is approximated as [41]

$$\mu_c = \hbar v_f \sqrt{\frac{\pi c_{ox} V_g}{e}} \qquad (8)$$

where $C_{ox}$ is the electrostatic gate capacitance, and $v_f$ is the Fermi velocity. The applied voltage $V_g$ between the silicon and the Graphene layer controls the carrier concentration $n \approx C_{ox}V_g/e$ [42]. More charge is induced on the surface of Graphene when applied external voltage increases, which in turn raises the chemical potential. The value of the gate voltage, carrier concentration, electric field, and chemical potential for Graphene nanoantenna is listed in Table 2.

The frequency reconfigurable behavior of the Graphene nanoantenna is achievable by varying the chemical potential in the range. The resonant frequency is reconfigured in a wide frequency range from 2.5 to 5.0 THz via the increase of chemical potential from 0.3 eV to 0.6 eV by the application of gate voltage 7.6 - 30.6 V. Furthermore, the increased mobility of the Graphene impacts on antenna performance. But, the decrease in the mobility of Graphene does not affect the resonant frequency, whereas return loss improves with increased mobility. Furthermore, it leads to the increase of the antenna radiation efficiency.

Table 2: The Performance enhancement of Graphene nanoantenna

| $V_g$ (V) | N (cm$^{-2}$) | E (MV/cm) | $\mu_c$ (eV) | $f_r$ (THz) | D (dBi) |
|---|---|---|---|---|---|
| 7.6 | $6.7\times10^{12}$ | 3.06 | 0.3 | 2.5 | 2.99 |
| 13.6 | $12\times10^{12}$ | 5.44 | 0.4 | 3.4 | 4.12 |
| 21.2 | $18.8\times10^{12}$ | 8.5 | 0.5 | 4.2 | 4.58 |
| 30.6 | $27\times10^{12}$ | 12.2 | 0.6 | 5.0 | 5.56 |

Graphene nanoantenna achieves a significant performance in the radiation pattern. The directivity increases and the back radiation decreases with the increase of Graphene chemical potential. Hence, the chemical potential of Graphene impacts the resonant frequency and radiation characteristics of the Graphene nanoantenna in the THz band. The directivity enhancement and frequency reconfiguration of the Graphene nanoantenna based on the chemical potential of Graphene is listed in Table. 2. Another significant performance of the Graphene nanoantenna is its stable impedance behavior upon reconfiguration. The Impedance value remains constant at all reconfigurable frequencies.

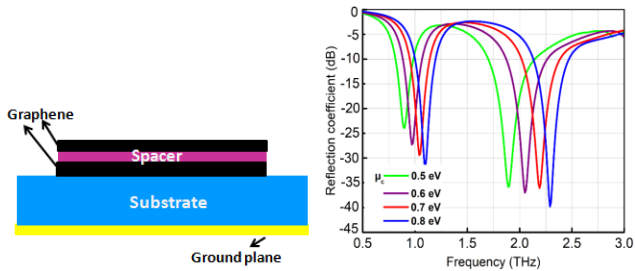

Figure 3: (a) Schematic of bilayer Graphene nanoantenna and its (b) dual-band reconfiguration.

The concept of the single-layer Graphene antenna can be extended to the bilayer Graphene antenna for dual-band operation. Bilayer Graphene nanoantenna provides easily dual-band reconfiguration and constant impedance characteristics upon reconfiguration at THz [27]. In bilayer Graphene nanoantenna, a two coupled Graphene layer is separated by a spacer, as shown in Fig. 3(a). The intermediate spacer provides a convenient way to control the conductivity of both the upper and lower Graphene layer using the electric field effect. The dual-band frequency reconfiguration performance of bilayer Graphene nanoantenna is shown in Fig. 3(b). The monolayer and bilayer Graphene provides frequency reconfigurability around one frequency and dual frequencies respectively, in the Graphene nanoantenna. The nanoantennas using monolayer and bilayer Graphene have the potential to be used for THz communication systems.

## 6 Graphene THz Hypersurface for Smart wireless Environment

THz communications are susceptible to acute path loss effects, owed to the standard power dissipation within space and molecular absorption [43]. Moreover, the Doppler Effect becomes more acute in such bands, resulting in problematic communications even at pedestrian speeds [43]. Thus, to attain its high promise for astounding data transfer rates, the THz band is likely to require the collaboration of the user device with a smart propagation environment.

Recently, Hypersurface has been proposed as a smart environment that can sense wavefronts emitted by user devices and manipulate their propagation. The environment routes emitted waves from the user devices to their end-destinations, fulfilling custom performance objectives such as shortest-distance propagation, interference minimization, eavesdropping cancellation and Doppler Effect mitigation (by ensuring that last bounce of the propagation is perpendicular to the trajectory of a mobile user) [44], [45]. The user-side Graphene nanoantennas and environment-side Hypersurfaces can build a promising smart THz wireless environment. HSF empowered THz wireless environment can optimize the propagation factor between THz wireless devices.

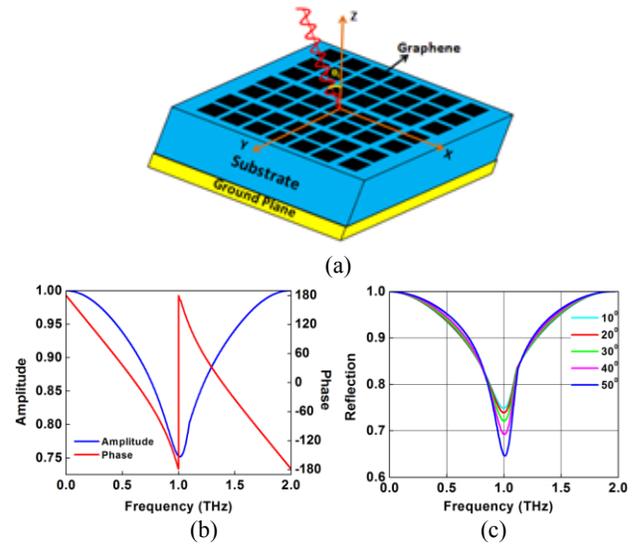

Figure 4: (a) Schematic of the Graphene MSF, (b) Amplitude and phase of reflection coefficient at normal incidence, and (c) Reflection at oblique incidence.

The strong optical response and plasmonic properties of Graphene enable novel metasurface at THz [46]. Owing to its tunable characteristics, Graphene is a promising candidate for THz HSF. The Graphene HSF structure is composed of an array of Graphene elements deposited on the silicon substrate and backed by a ground plane, as shown in Fig. 4(a). Fig. 4(b) shows the reflection amplitude and phase of Graphene HSF. Graphene THz HSF provides around 75% efficiency of reflection at normal angle of incidence at 1 THz, and this has the capability of $360^0$ reflection phase variation. Reflection efficiency of Graphene HSF can further be increased by controlling the properties of Graphene. Graphene HSF shows around 65 to 75% of reflection efficiency under oblique angle incidences from 0° to 50°, as shown in Fig. 4(c). The perfect absorption and frequency reconfiguration is also easily achieved in Graphene THz HSF [46]. It reveals that Graphene HSF has the potential to be used for the smart THz wireless environment.

## 7 Conclusions

THz nanoantennas play a significant role in achieving better performance in THz wireless communications. This paper firstly discussed the suitable materials for the nanoantenna design at THz. Then it discussed the performance enhancements offered by nanoantennas.

In this paper, the performance of metal copper, carbon nanotube (CNT), and Graphene nanoantennas are compared in order to find the best material for nanoantenna at THz. Material properties and its electromagnetic behavior at THz are analyzed to justify the of THz nanoantenna performance. A critical comparison of the performance of copper, carbon nanotube, and Graphene shows that the Graphene nanoantennas have better performance in terms of directivity and miniaturization. Further, this paper presented the performance enhancement of Graphene nanoantenna in terms of high miniaturization, high directivity, low reflection, low back lobe radiation, frequency reconfiguration, and stable impedance by controlling the graphene surface conductivity. The performances of Graphene nanoantenna make it a promising candidate for THz communication. This paper finally discussed THz Hypersurfaces using Graphene meta-atoms. The user-side Graphene antennas and environment-side Hypersurfaces can build a promising smart THz wireless environment.

## ACKNOWLEDGMENT

This research is supported by the European Union via the Horizon 2020: Future Emerging Topics - Research and Innovation Action call (FETOPEN-RIA), grant EU736876, Project VISORSURF (http://www.visorsurf.eu).

## REFERENCES

[1] A.J.Paulraj, D.A.Gore, R.U.Nabar, and H.Bolcskei, "An overview of MIMO communications—A key to gigabit wireless," Proc. IEEE, vol. 92, pp. 198–218, 2004.
[2] I. F. Akyildiz, J. M. Jornet, and C. Han, "Terahertz band: Next frontier for wireless communications," Physical Communication, vol.12, pp. 16-32, 2014.
[3] J. Federici and L. Moeller, "Review of terahertz and sub terahertz wireless communications", Journal of Applied Physics, vol. 107, pp.111101, 2010.
[4] I. Malhotra, K. R. Jha, G.Singh, "Analysis of highly directive photoconductive dipole antenna at terahertz frequency for sensing and imaging applications," Optics Communications, vol. 397, pp. 129-139, 2017.
[5] K. Konstantinidis, A. P. Feresidis, Y. Tian, X. Shang, and M. J. Lancaster, "Micromachined terahertz Fabry–Perot cavity highly directive antennas," IET Microw., Antennas Propag., vol. 9, no. 13, pp. 1436–1443, 2015.
[6] A. J. Alazemi, H. H. Yang, and G. M. Rebeiz, "Double bow-tie slot antennas for wideband millimeter-wave and terahertz applications," IEEE Trans. THz Sci. Technol., vol. 6, no.5, pp. 682–689, 2016.
[7] H. Yu ; J. Yu, Y. Yao ; X. Liu ; X. Chen, "Wideband circularly polarised horn antenna with large aspect ratio for terahertz applications, Electronics Letters, vol. 56, no.1, pp. 11-13, 2020
[8] T. Niu, W. Withayachumnankul, B. S.Y. Ung, H. Menekse, M. Bhaskaran, S. Sriram, C. Fumeaux, "Experimental demonstration of reflectarray antennas at terahertz frequencies," Opt. Expr., vol. 21, no. 3, pp. 2875-2889, 2013.
[9] Z. Hao , J. Wang , Q. Yuan ; W. Hong, "Development of a Low-Cost THz Metallic Lens Antenna', IEEE Antennas and Wireless Propagation Letters, vol.16, pp. 1751-1754, 2017.
[10] G. W. Hanson, "Current on an infinitely-long carbon nanotube antenna excited by a gap generator," IEEE Trans. Antennas Propag., pp. 76–81, 2006.
[11] M. Zhao, M. Yu, and H. Robert Blick, "Wavenumber-domain theory of terahertz single-walled carbon nanotube antenna," IEEE Journal of Selected Topics in Quantum Electronics, vol. 18, no. 1, pp. 166–175, 2012.
[12] G. W. Hanson, "Fundamental transmitting properties of carbon nanotube antennas," IEEE Trans. Antennas Propag., vol. 53, pp.3426–3435, 2005.
[13] A. M. Attiya, "Lower frequency limit of carbon nanotube antenna," Prog. Electromagn. Res., vol. 94, pp. 419–433, 2009.
[14] Y. Huang, W.-Y. Yin, and Q. H. Liu, "Performance prediction of carbon nanotube bundle dipole antennas," IEEE Trans. Nanotechnol., vol. 7, no. 3, pp. 331–337, 2008.
[15] S. F. Mahmoud and A. R. Alajmi, "Characteristics of a new carbon nanotube antenna structure with enhanced radiation in the sub-terahertz range," IEEE Trans. Nanotechnol., vol. 11, no. 3, pp. 640–646, 2012.
[16] A. R. Alajmi and S. F. Mahmoud, "Investigation of multiwall carbon nanotubes as antennas in the sub terahertz range," IEEE Trans. Nanotechnol, vol. 13, no. 2, pp. 268-273, 2014.
[17] S. Dash, and A. Patnaik, "Material selection for THz antennas", Microwave and Optical Technology Letters, vol. 60, pp. 1183-1187, 2018.
[18] S. Dash, and A. Patnaik, "Performance of Graphene Plasmonic Antenna in Comparison with their Counterparts for Low-Terahertz Applications", Plasmonics, Vol. 13, no. 6, pp. 2353-2360, 2018.
[19] Josep Miquel Jornet and Ian F. Akyildiz, "Graphene-based Plasmonic Nano-Antenna for Terahertz Band Communication in Nanonetworks," IEEE Journal on Selected Areas in Communications Part 2, vol. 31, No. 12, pp. 685-694, 2013.
[20] S. Dash, A. Patnaik and B. K Kaushik, "Performance Enhancement of Graphene Plasmonic Nanoantenna For THz Communication", IET Microwaves, Antennas & Propagation, Vol. 13, no. 1, pp. 71-75, Jan 2019.
[21] S. Dash, and A. Patnaik, "Sub-wavelength Graphene Planar nanoantenna for THz Application", Materials Today: Proceedings, vol. 18, Part 3, pp. 1336-1341, 2019.
[22] S. Prakash, S. Dash and A. Patnaik, "Reconfigurable Circular Patch THz Antenna using graphene stack based SIW Technique", 2018 IEEE Indian Conference on Antennas and Propogation, Hyderabad, India, pp. 1-3, 2018.
[23] E. Carrasco and J. Perruisseau-Carrier, "Reflectarray Antenna at Terahertz Using Graphene", IEEE Antennas and Wireless Propagation Letters, vol. 12, pp. 253-256, 2013.
[24] Z. Xu, X. Dong and J. Bornemann, "Design of a Reconfigurable MIMO System for THz Communications Based on Graphene Antennas," IEEE Transactions on Terahertz Science and Technology, vol. 4, pp. 609-617, 2014.
[25] S. Dash, and A. Patnaik, "Graphene Plasmonic Bowtie Antenna for UWB THz Application", IEEE 24th National Conference on Communications, Hyderabad, India, pp. 1-4, 2018. DOI: 10.1109/NCC.2018.8599940
[26] M. Esquius-Morote, J. S. Gómez-Díaz, and J. Perruisseau-Carrier, "Sinusoidally-Modulated Graphene Leaky-Wave Antenna for Electronic Beam scanning at THz," IEEE Transactions on Terahertz Science and Technology, vol. 4, pp. 116-122, 2014.
[27] S. Dash, and A. Patnaik, "Dual-Band Reconfigurable Plasmonic Antenna using Bilayer Graphene", Proc. of IEEE International Symposium on Antennas and Propagation AP-S 2017, San Diego, California, USA, pp. 921-922, July 9-14, 2017.
[28] R. Piesiewicz, T. Kleine-Ostmann, N. Krumbholz, D. Mittleman, M. Koch, J. Schoebel, and T. Kurner, "Short-range ultra-broadband terahertz communications: Concepts and perspectives," IEEE Antennas Propag. Mag., vol. 49, no. 6, pp. 24–39, 2007.
[29] ] K. C. Huang, and Z. Wang, Terahertz terabit wireless communication, IEEE Microwave Magazine, vol. 12(4), pp. 108–116, 2011.
[30] I. Akyildiz, J. Jornet, and C. Han, "TeraNets: Ultra-broadband communication networks in the terahertz band," IEEE Wireless Commun., vol. 21, no. 4, pp. 130–135, Aug. 2014.
[31] J. M. Jornet and I. F. Akyildiz, "Channel modeling and capacity analysis for electromagnetic wireless nanonetworks in the terahertz band," IEEE Trans. Wireless Commun., vol. 10, no. 10, pp. 3211–3221, Oct. 2011
[32] K. S. Novoselov *et. al*., Girgorieva, and A. A. Firsov, "Electric field effect in atomically thin carbon films," Science, vol. 306, pp. 666-669, 2004.
[33] K. I. Bolotin *et. al*. , "Ultrahigh electron mobility in suspended graphene," Solid State Commun., vol. 146, pp. 351–355, 2008.
[34] J. Yu, G. Liu, A. V. Sumant, V. Goyal, A. A. Balandin, "Graphene-on-diamond devices with increased current-carrying capacity: carbon sp2-on-sp3 technology," Nano letters, vol. 12, pp. 1603–1608, 2012.
[35] F. H. L. Koppen *et. al*., "Graphene plasmonics: A platform for strong light matter interactions," Nano Lett., vol.11 (8), pp. 3370-3377, 2011.
[36] M. Jablan, H. Buljan, and M. Soljacic, "Plasmonics in graphene at infrared frequencies," Physical Review B, vol. 80, pp. 245435(1-7), 2009.
[37] V. Gusynin, S. Sharapov, and J. Carbotte, "Magneto-optical conductivity in graphene," J. Phys.: Condens. Matter, vol. 19, pp. 026222(1–28), 2006.
[38] I. ijima, "Helical microtubules of graphitic carbon," Nature, vol. 354, pp. 56–58, 1991.
[39] P. L McEuen, M. S. Fuhrer, and H. K Park, "Single-walled carbon nanotube electronics," IEEE Trans. Nanotechnol., vol. 1, pp. 78-85, 2002.
[40] C. Kittel, Introduction to solid state physics, 6th Ed. Wiley, New York, 1986.
[41] C. Zeng, X. Liu, and G. Wang, "Electrically tunable graphene plasmonic quasicrystal metasurfaces for transformation optics," Sci. Rep., vol. 4, no. 5763, pp. 1-8, 2014.
[42] P. Y. Chen, and A. Alu, "Atomically Thin Surface Cloak Using Graphene Monolayers," ACS Nano, vol. 5, no. 7, pp. 5855-5863, 2011.
[43] I. F. Akyildiz, S. Nie, S.-C. Lin, and M. Chandrasekaran, 5g roadmap: 10 key enabling technologies, Computer Networks, vol. 106, pp. 1748, 2016.
[44] C. Liaskos, A. Tsioliaridou, S. Nie, A. Pitsillides, S. Ioannidis, and I. Akyildiz, Modeling, simulating and conguring programmable wire-less environments for multi-user multi-objective networking, CoRR, vol. abs/1812.11429, 2018.
[45] C. Liaskos, N. Shuai, A. Tsioliaridou, A. Pitsillides, S. Ioannidis, and I. Aky- ildiz, A novel communication paradigm for high capacity and security via programmable indoor wireless environments in next generation wireless sys- tems, Ad Hoc Networks, vol. 87, pp. 116, may 2019.
[46] S. Dash, C. Liaskos, I. F. Akyildiz, . Pitsillides, "Wideband Perfect Absorption Polarization Insensitive Reconfigurable Graphene Metasurface for THz Wireless Environment", MTTW'19, IEEE Workshop on microwave Theory and Techniques in Wireless Communication, 1 - 2 Oct 2019, Riga, Latvia. DOI: 10.1109/MTTW.2019.8897231.